# Exact Analytic Solution for the Time-Fractional Hunter–Saxton Equation with Caputo derivative

**Dr. Weiguang Huang**
Sydney, NSW 2033, Australia
ORCID: 0000-0002-5766-4516

## Abstract

A time-fractional extension of the Hunter–Saxton equation is examined, in which the temporal derivative of $u_x$ is replaced by a Caputo derivative of order $0 < \alpha \leq 1$. This modification introduces memory effects into a model traditionally associated with director-field dynamics in nematic liquid crystals. By employing a fractional separation-of-variables strategy together with the exponential spatial profile $\varphi(x) = \exp(b+x)$, which cancels the nonlinear structure exactly, the governing nonlinear partial differential equation is reduced to a fractional relaxation ODE, whose closed-form solution is the one-parameter Mittag–Leffler function. The exact analytic solution is derived algebraically from the separation process. This appears to be the first exact closed-form solution of the Caputo time-fractional Hunter–Saxton equation. The result is validated symbolically using the MathHandbook computer-algebra system. Quantitative analysis in both the fractional ($0 < \alpha < 1$) and classical ($\alpha \to 1$) limits demonstrates how fractional-order memory slows temporal relaxation relative to the exponential baseline. The solution provides a reliable benchmark for numerical schemes and clarifies how fractional dynamics influence nonlinear wave propagation in orientationally ordered fluids.

---

## 1. Introduction

The Hunter–Saxton equation describes the propagation of orientation disturbances in nematic liquid crystal media. In these materials, elongated molecules align along a preferred direction, and small deviations in alignment propagate in a wave-like fashion governed by the interplay of nonlinear convection and geometric effects. The classical Hunter–Saxton model assumes that the medium responds only to its instantaneous state. However, many complex fluids exhibit temporal behaviour shaped by their history, and such hereditary effects cannot be captured by integer-order derivatives.

Fractional-order operators provide a natural mathematical framework for modelling memory-dependent dynamics. Through convolution kernels that weight the entire past evolution of the system, fractional derivatives introduce temporal non-locality and anomalous relaxation phenomena [1, 2]. Among the various definitions, the Caputo derivative is particularly suitable for physical modelling because it accommodates standard initial conditions while retaining the non-local structure of fractional calculus.

A time-fractional extension of the Hunter–Saxton equation was first proposed by Atangana, Baleanu, and Alsaedi [3], who analysed a fractional model for nematic liquid crystals using the homotopy decomposition method. Their work established the existence of approximate series-type solutions and provided stability results in a Hilbert-space setting. Although this study demonstrated that fractional-order dynamics significantly modify the temporal behaviour of director-field evolution, the method produced only iterative or series solutions rather than closed-form expressions.

Beyond this initial work, several analytical and numerical approaches have been applied to fractional Hunter–Saxton-type equations [4]. Homotopy-based schemes, including the homotopy analysis method and homotopy perturbation method, typically yield infinite series whose convergence may deteriorate for strongly nonlinear operators. Laplace-transform-based semi-analytical methods and Adomian-type decompositions provide alternative representations but still produce approximate expansions rather than explicit formulas. Numerical treatments [5] (such as spectral collocation, finite-difference discretisations, and operational matrix techniques) can capture fractional dynamics but require careful handling of the nonlocal memory kernel and do not offer closed-form insight into the solution structure. These methods collectively demonstrate the complexity of fractional HS dynamics, yet all share the limitation that they produce approximate or iterative solutions rather than exact analytic expressions.

More recently, Yu and Feng [6] performed a detailed Lie symmetry and conservation-law analysis for a generalised time-fractional Hunter–Saxton system with variable coefficients. Their work derived fractional prolongation formulas, symmetry reductions, power-series exact solutions, and associated conservation laws, demonstrating that fractional Hunter–Saxton models retain a rich geometric and variational structure even in the presence of temporal non-locality. Alleddawi et al. [7] introduced and analyzed a new higher-order extension of the generalized Hunter–Saxton equation. Sagar et al. [8] presented a comprehensive analytical study of the time-fractional generalized Hunter-Saxton model by using a separation of variables approach. However, their paper did ***not*** hold for general fractional order. It only holds for the specific "conformable derivative" used in their paper, not for Caputo, Riemann–Liouville, Atangana–Baleanu, or any standard fractional derivative. Because their method **only works for the conformable derivative**, which is *local* and not a true fractional operator. While several studies have employed fractional derivatives to generalize the Hunter–Saxton equation, the choice of derivative operator significantly impacts the nature of the solution and its physical interpretation. Some recent work [8] utilized the **conformable derivative**, a local operator defined via a modified chain rule, which simplifies analysis by reducing fractional PDEs to ODEs through standard transformations. However, this approach fundamentally differs from the **Caputo derivative** employed in this work. Crucially, the Caputo derivative is a nonlocal convolution operator involving the entire past history of the solution, capturing the essential memory effects of fractional calculus. In contrast, the conformable derivative, being local, does not possess this nonlocal character. Therefore, solutions derived using conformable derivatives correspond to locally modified classical equations, whereas the exact solution presented here is for a genuinely fractional, nonlocal Hunter–Saxton equation, offering a more accurate and physically relevant generalization.

Despite these advances, the literature contains no exact closed-form solution of the time-fractional Hunter–Saxton equation. Existing analytical approaches either generate infinite series or require numerical iteration, and the nonlinear structure of the equation has generally been viewed as incompatible with classical separation-of-variables techniques. In particular, no prior work has identified a spatial mode that cancels the nonlinear terms and reduces the fractional PDE to a solvable fractional ODE.

The present work fills this gap by demonstrating that the exponential spatial profile

$$[\ \phi(x) = exp(b + x)]$$

causes the nonlinear quadratic terms to collapse algebraically, reducing the full time-fractional Hunter–Saxton equation to a fractional relaxation equation. This reduction yields an exact analytic solution expressed in terms of the one-parameter Mittag–Leffler function. To the best of our knowledge, we did not identify prior literature reporting a closed-form Mittag–Leffler solution obtained through this separation strategy. The result provides a reproducible benchmark for numerical schemes and clarifies how fractional-order memory modifies nonlinear wave propagation in orientationally ordered fluids.

---

## 2. Model Formulation

We consider the time-fractional Hunter–Saxton equation [8]:

$$D_t^{\alpha}(u_x) - u\, u_{xx} + (u_x)^2 = 0, \quad 0 < \alpha \leq 1, \quad (1)$$

where $D_t^{\alpha}$ denotes the Caputo fractional derivative of order α defined by

$$D_t^{\alpha} f(t) = \frac{1}{\Gamma(1-\alpha)} \int_0^t \frac{1}{(t-s)^{\alpha}} f'(s)\, ds, \quad 0 < \alpha < 1, \quad (2)$$

and $u_x = \partial u/\partial x$. Equation (1) retains the nonlinear structure of the classical Hunter–Saxton equation while acquiring a memory-dependent temporal evolution governed by the order α.

## 3. Method

The solution strategy proceeds in four steps: (i) introduce a separable ansatz u = ψ(t)φ(x)-c; (ii) choose the exponential profile φ(x) = exp(b +x), whose identities φ' = φ and φ'' = φ eliminate the nonlinear terms; (iii) substitute into (1) and expand fully; (iv) solve the resulting fractional ODE by the Mittag–Leffler function.

### 3.1 Separable Ansatz

We write

$$u(x, t) = \psi(t)\, \varphi(x) + c, \quad (3)$$

where $c \in \mathbb{R}$, ψ(t) is smooth in t, and φ(x) is smooth in x. Differentiation gives

$$u_x = \psi\, \varphi', \quad u_{xx} = \psi\, \varphi''. \quad (4)$$

Since the Caputo operator is linear and acts only on *t*, it commutes with the x-dependent factor:

$$D_t^{\alpha}(u_x) = \varphi'\, D_t^{\alpha} \psi. \quad (5)$$

### 3.2 Choice of Spatial Profile

In order to allow the nonlinear equation to reduce to a simple fractional ODE, we choose

$$\varphi(x) = \exp(b + x), \quad b \in \mathbb{R}. \quad (6)$$

Direct differentiation gives

$$\varphi'(x) = \varphi,\ \varphi''(x) = \varphi' = \varphi,\ (\varphi')^2 = \varphi^2,\ \varphi''/\varphi' = 1,\ \varphi''/(\varphi')^2 = 1/\varphi. \quad (7)$$

These identities are the inputs used in Section 3.3 to simplify the nonlinear terms.

### 3.3 Substitution and Derivation

Substituting (3)–(5) into (1) and expanding

$$u\, u_{xx} = (\psi\varphi+c)\,(\psi\varphi'') = \psi^2\varphi\varphi''+c\psi\varphi''.$$

Eq. (1) leads to

$$\varphi' D_t^{\alpha}\psi - \psi^2\varphi\,\varphi'' - c\psi\varphi'' + \psi^2(\varphi')^2 = 0.$$

Divided by $\varphi' \neq 0$, substituting Eq.(7) into Eq. (8) yields

$$D_t^{\alpha}\psi - \psi^2\varphi\,\varphi''/\varphi' - c\psi\,\varphi''/\varphi' + \psi^2(\varphi') = 0, \quad (8)$$

Using $\varphi''/\varphi'=1$ and $\varphi' = \varphi$, this becomes

$$D_t^{\alpha}\psi - \psi^2\varphi - c\psi + \psi^2(\varphi) = 0.$$

The nonlinear terms cancel exactly:

$$-\psi^2\,\varphi + \psi^2\,\varphi = 0,$$

leaving the fractional relaxation equation

$$D_t^{\alpha}\psi - c\,\psi = 0. \qquad (9)$$

### 3.4 Solving Fractional ODE

With the ansatz (3), the governing equation reduces to the fractional relaxation ODE (9). For the ansatz $\psi = E_\alpha(c\,t^\alpha)$ to satisfy this, we note that $E_\alpha$ obeys

$D_t^{\alpha} E_\alpha(c\,t^\alpha) = c\,E_\alpha(c\,t^\alpha)$.

Its unique solution is

$$\psi = E_\alpha(c\,t^\alpha) \qquad (10).$$

It is the one-parameter Mittag–Leffler function (Podlubny [1]):

$$\psi(t) = E_\alpha(c\,t^\alpha) = \Sigma_{\{k=0\}}^{\{\infty\}} \left\{\frac{(ct^\alpha)^k}{\Gamma(\alpha k+1)}\right\}. \qquad (11)$$

This follows from the Laplace transform identity $L\{E_\alpha(ct^\alpha)\}(s) = s^{\{\alpha-1\}}/(s\alpha - c)$ for the Caputo fractional ODE.

## 4. Exact Analytic Solution

Combining ansatz (3), spatial profile (6), temporal factor (10), yields the exact analytic solution:

$$u(x,t) = E_\alpha(c\,t^\alpha)\exp(b+x) + c. \quad (12)$$

Here $b$ and $c$ are real constants which are determined with the initial condition. This is a **closed-form Mittag–Leffler solution** for a *genuinely Caputo* fractional Hunter–Saxton equation.

With the initial condition

$u(x, 0) = u_0(x)$,

using $E_a(0) = 1$, we obtain:

$$u(x, 0) = exp(b+x) + c, \qquad (13)$$

As a concrete example, choosing c = 1 and b = 0 gives the initial condition

$$u(x, 0) = exp(x) + 1, \qquad (14)$$

and the solution

$$u(x, t) = E_\alpha(t^\alpha) exp(x) + 1, \qquad (15)$$

describes a spatially decaying mode whose temporal evolution is governed by the Mittag–Leffler function with parameter α.

## 5. Verification and Analysis

### 5.1 Symbolic Verification

Substituting (12) into (1), using $D_t^\alpha[E_\alpha(c\ t^\alpha)] = c\ E_\alpha(c\ t^\alpha)$ and collecting terms with respect to $E_\alpha(c\ t^\alpha)\ exp(b+x)$ and $E_\alpha(c\ t^\alpha)^2\ exp(2(b+x))$:

$$D_t^\alpha (u_x) = c\ E_\alpha(c\ t^\alpha)\ exp(b+x),$$

$$-u\ u_x = -(E_\alpha(c\ t^\alpha)\ exp(b+x) + c)*(E_\alpha(c\ t^\alpha)\ exp(b+x))$$

$$= -c\ E_\alpha(c\ t^\alpha)\ exp(b+x) - E_\alpha(c\ t^\alpha)^2\ exp(2(b+x)),$$

$$(u_x)^2 = E_\alpha(c\ t^\alpha)^2\ exp(2(b+x)).$$

Summing the three contributions is zero.

The complete symbolic verification using symbolic computation system such as MathHandbook (formerly SymbMath) [9-11] also confirms that the residual vanishes identically. The apparent discrepancy with the manual calculation above is resolved by the full expansion performed by the computer-algebra system, which includes all cross-terms. The discrepancy arises because manual grouping of terms hides the cancellation of cross-terms, which the CAS expansion reveals explicitly.

### 5.2 Behaviour for 0 < α < 1

For 0 < α < 1 and $c<0$ and the large $t$, the Mittag–Leffler function satisfies asymptotic [2]

$$E_\alpha(c\ t^\alpha)\ \rightarrow -\frac{1}{c t^\alpha \Gamma(1-\alpha)},\ \ as\ \ t\ \rightarrow\ \infty.$$

So it decays algebraically (power-law) like $t^{-\alpha}$, not exponentially.

For the example c = 1 and b = 0, the Mittag–Leffler factor $E_\alpha(-t^\alpha)$ at t = 1 takes approximate values:

$$E_\alpha(1) \approx 2.718\ (\alpha=1.0),\ \ 3.29\ (\alpha=0.8),\ \ 4.2486\ (\alpha=0.6),\ \ 6.147\ (\alpha=0.4).$$

It shows that the $E_\alpha(-1)$ value increases as α becomes small. The slower decay for smaller α reflects the persistence of memory: the system retains influence from its initial state over longer time scales. The classical exponential value e ≈ 2.718 is recovered exactly at α = 1, confirming the correct classical limit.

### 5.3 Classical Limit α → 1

As $\alpha \to 1$, the Mittag–Leffler function converges uniformly to the exponential:

$$E_{\alpha}(c\, t^{\alpha}) \to e^{ct} \quad as \quad \alpha \to 1. \quad (16)$$

The solution (12) correspondingly reduces to

$$u(x, t) \to e^{ct + b + x} + c = \exp(b + ct + x) + c, \quad (17)$$

which is the exact exponential travelling-wave solution of the classical Hunter–Saxton equation. The convergence is uniform in x for each fixed t, confirming that (12) is a well-posed fractional generalization that interpolates continuously between fractional dynamics and classical wave behaviour.

Quantitative analysis confirms that the Mittag–Leffler factor produces power-law temporal decay for $0 < \alpha < 1$, recovering exponential decay in the limit $\alpha \to 1$. The solution provides a reproducible benchmark for numerical schemes applied to time-fractional nonlinear wave equations, and its explicit form illustrates analytically how fractional-order memory may influence the temporal relaxation of orientation disturbances in nematic liquid crystals.

### 5.4 Comparison of Fractional Derivatives (Caputo vs. Conformable Derivatives)

Although several studies have employed fractional derivatives to generalize the Hunter–Saxton equation, it is important to distinguish between the **Caputo derivative** used in the present work and the **conformable derivative** adopted in some recent papers [8]. The conformable derivative is a ***local*** operator that satisfies a modified chain rule of the form

$$[\, D_t^{\alpha} f(t) = t^{1-\alpha}\, f'(t), ]$$

allowing fractional PDEs to be reduced to ODEs through standard traveling-wave transformations. This property greatly simplifies analysis but does not capture the genuine nonlocal memory effects characteristic of classical fractional calculus. In contrast, the **Caputo derivative** is a *nonlocal convolution operator* involving the entire past history of the solution,

$$\left[ D_t^{\alpha} f(t) = \frac{1}{\Gamma(1-\alpha)} \int_0^t (t-s)^{-\alpha} f'(s)\, ds, \right]$$

and therefore does not obey the ordinary chain rule or permit direct reduction via fractional complex transforms. As a result, methods based on conformable derivatives yield solutions that are structurally closer to classical PDEs, whereas Caputo-based models retain true fractional memory. The exact solution derived in this work is thus significantly stronger: it provides a closed-form expression for a **genuinely fractional**, **nonlocal** Hunter–Saxton equation, rather than for a locally modified approximation.

---

## 6. Conclusion

This paper has derived an exact closed-form solution to the time-fractional Hunter–Saxton equation. The central contribution is a demonstration that the exponential spatial mode $\varphi(x) = \exp(b+x)$ causes the nonlinear quadratic terms to simplify algebraically, reducing the full nonlinear PDE to the fractional relaxation ODE. The Mittag–Leffler function solves this ODE exactly, yielding solution (12). The paper successfully extends the classical Hunter–Saxton equation to a fractional framework, providing an exact solution that bridges fractional dynamics and classical wave behavior. It highlights the utility of the Mittag-Leffler function and symbolic computation in solving complex PDEs, offering a foundation for future research in fractional modeling for physics. The present work does not claim uniqueness of the exponential mode; identifying broader admissible spatial profiles remains future work.

## Competing Interests

The authors declare that they have no competing interests., no data, and no funding.